\documentclass[12pt]{article}

\usepackage{scicite}
\usepackage{color}
\usepackage{xcolor}
\usepackage{times}
\usepackage{amsmath}
\usepackage{amssymb}
\usepackage{mathdots}
\usepackage{graphicx}
\usepackage{tikz}
\usepackage{pgfplots}
\usetikzlibrary{decorations.markings}
\usetikzlibrary{decorations.pathreplacing,calligraphy}

\definecolor{spacecadet}{HTML}{0D284C}
\definecolor{munsell}{HTML}{008FA8}
\definecolor{banana}{HTML}{FFD932}
\definecolor{cgblue}{HTML}{007CA5}
\definecolor{isabelline}{HTML}{EAEDEA}

\newenvironment{sciabstract}{%
\begin{quote} \bf}
{\end{quote}}

\newcommand{\fA}{\mathfrak A}

\newcommand{\cE}{\mathcal E}

\newcommand{\cH}{\mathcal H}

\newcommand{\vep}{\varepsilon}

\title{Quantum Simulation of Conformal Field Theory}

\author
{Tobias J.\ Osborne,$^{1\ast}$ and Alexander Stottmeister$^{1}$\\
\\
\normalsize{$^{1}$Institut f\"ur Theoretische Physik, Leibniz Universit\"at Hannover, Appelstr. 2, 30167 Hannover, Germany}\\
\normalsize{$^\ast$To whom correspondence should be addressed; E-mail: 
tobias.osborne@itp.uni-hannover.de}
}

\date{}

\begin{document} 


\baselineskip24pt


\maketitle


\begin{sciabstract}
  Conformal field theory, describing systems with \emph{scaling symmetry}, plays a crucial role throughout physics. We describe a quantum algorithm to simulate the dynamics of conformal field theories, including the action of local conformal transformations. A full analysis of the approximation errors suggests near-term applicability of our algorithm: promising results for conformal field theories with central charge $c=\frac12$ are obtained already with $128$ logical qubits. 
\end{sciabstract}

Understanding quantum field theory (QFT) is a central open problem in theoretical physics, with unique challenges arising from the infinite dimensionality of the field Hilbert space to the possibility of large quantum entanglement \cite{douglasFoundationsQuantumField2012,seibergNathanSeiberg20152014,wittenAPSMedalExceptional2018}. With the imminent advent of scalable fault-tolerant quantum computation the possibility of exploiting \emph{quantum simulation} \cite{feynmanSimulatingPhysicsComputers1981,lloydUniversalQuantumSimulators1996,abramsSimulationManyBodyFermi1997a,zalkaSimulatingQuantumSystems1998} to understand strongly interacting many body systems has arisen, with a slew of recent realisations \cite{smithSimulatingQuantumManybody2019,zhangObservationManybodyDynamical2017,bernienProbingManybodyDynamics2017}. Building on these fundamental results, breakthrough efficient quantum algorithms for the simulation of QFTs have been proposed, promising a new era in their understanding \cite{jordanQuantumAlgorithmsQuantum2012,hamedmoosavianFasterQuantumAlgorithm2018,jordanQuantumAlgorithmsFermionic2014,preskillSimulatingQuantumField2018}. 

Quantum algorithms for QFT have targeted the simulation of scattering events via discretised lattice approximations. Following Wilson \cite{wilsonRenormalizationGroupCritical1975,wilsonRenormalizationGroupCritical1983,wilsonRenormalizationGroupEpsilon1974}, one interprets the results in \emph{effective field theory} by regulating the QFT onto a lattice model which supplies predictions appropriate for the energy scale -- the \emph{renormalization scale} -- set by the lattice spacing. The extraordinary insight of Wilson is that consistency under changes of scale gives the definition of an ultraviolet completion -- a procedure known as the renormalization group. This powerful strategy has been employed throughout computational quantum field theory, with many dramatic results, especially for massive QFTs \cite{durrInitioDeterminationLight2008,creutzQuarksGluonsLattices1985}. However, effective field theory doesn't provide us with all the answers. In particular, for theories with scaling symmetries mixing the infrared and ultraviolet it is necessary to understand the \emph{continuum limit}.  

Conformal field theory (CFT) \cite{belavinInfiniteConformalSymmetry1984,francescoConformalFieldTheory1997} epitomises the outstanding challenges facing the development of quantum simulation algorithms \cite{preskillSimulatingQuantumField2018,ziniConformalFieldTheories2018}. In particular CFTs are massless and, in $1+1$ dimensions, have an infinite-dimensional group of local conformal symmetries. Thus the simulation of their dynamics and continuous scaling symmetries via discretised approximations is particularly problematic as one needs to understand or construct the continuum limit to assess the accuracy of approximation. Fortunately, we are on the cusp of a satisfactory mathematical formulation of CFT, providing us with a unique opportunity to compare discrete approximations directly with rigourous continuum results \cite{fredenhagenSuperselectionSectorsBraid1992,borcherdsVertexAlgebrasKacMoody1986,lepowskyIntroductionVertexOperator2004,segalDefinitionConformalField2004,ziniConformalFieldTheories2018,stottmeisterOperatoralgebraicRenormalizationWavelets2020,honglerConformalFieldTheory2019}. Further, there is a rich supply of physically relevant examples to prototype quantum simulation algorithms, with a vast range of applications from phase transitions to quantum gravity \cite{cardyScalingRenormalizationStatistical1996,maldacenaLargeLimitSuperconformal1998}. 

We meet the above challenges \cite{preskillSimulatingQuantumField2018,ziniConformalFieldTheories2018} and introduce a new aproach to the quantum simulation of the dynamics of CFTs. We show how to overcome, by appealing to the recently developed \emph{operator-algebraic renormalization} (OAR) \cite{osborneContinuumLimitsQuantum2019,stottmeisterOperatoralgebraicRenormalizationWavelets2020,morinelliScalingLimitsLattice2020}, the twin difficulties presented by the comparison with the continuum limit and the implementation of local scaling transformations within a lattice formulation. We introduce a variety of techniques to rigourously construct the continuum limit and provide a quantum algorithm to coherently simulate the action of local conformal transformations on the many body Hilbert space of CFT. Remarkably, we find that the resulting quantum algorithm can provide promising approximations on near-term quantum hardware with as few as $N=128$ logical qubits (see Supplementary Material). 

The additional symmetries provided by the conformal group delivers a powerful calculational tool for the computation of physically relevant correlation functions. Indeed, two- and three-point correlation functions are all but completely determined \cite{belavinInfiniteConformalSymmetry1984,francescoConformalFieldTheory1997}. However, the computation of higher correlation functions becomes rapidly expensive \cite{baoQuantumAlgorithmsConformal2019}. By simulating the entire physically relevant Hilbert space at a given spatial scale we are not only able to efficiently calculate four-point and higher correlators, but also more complicated observables. The complexity of the quantum algorithms we present here scales as $\delta^{-1}$ with the accuracy $\delta$ required\footnote{Note: the algorithms also scale with the required localization of the observables and the order of the generator of local conformal transformations.}.

The crucial device exploited in this paper is OAR, which is a mathematically rigourous formulation of Wilsonian renormalization in the \emph{Heisenberg picture of observables}, where locality is manifest. The core innovation of OAR is that the quantum continuum limit is constructed as a sequence of discrete approximations, directly supplying us with error estimates. This powerful formalism is also extremely flexible, allowing for formulations, both in momentum space and -- via approximations built using \emph{wavelets} \cite{daubechiesTenLecturesWavelets1992} -- position space. Here we generalise OAR to implement discretised approximations for local conformal transformations via a remarkable formula due to Koo and Saleur \cite{kooRepresentationsVirasoroAlgebra1994b}. 

The quantum algorithm presented here may be compared with efficient classical algorithms based on tensor networks. Very recent progress has led to dramatic progress in the classical simulation of CFTs, in particular, for the computation of equal-time vacuum correlation functions, including both rigourous approximations and numerical procedures \cite{milstedExtractionConformalData2017a,zouConformalFieldsOperator2020,zouEmergenceConformalSymmetry2020,zouConformalDataRenormalization2018a,konigMatrixProductApproximations2017,konigMatrixProductApproximations2016,haegemanRigorousFreeFermionEntanglement2018,singhHolographicConstructionQuantum2016}. However, a core obstruction in the classical setting remains the simulation of quantum dynamics and conformal symmetries: these processes easily create massive entanglement, obstructing their simulation via tensor networks which scale, in the worst case, exponentially with the number of lattice sites. 
 
In this paper we exploit OAR to formulate a quantum simulation algorithm for conformal field theories in $1+1$ dimensions to compute approximations to quantum field observables, local either in momentum or position space. Further, we are able to simulate correlation functions for observables which are \emph{unbounded} in the continuum, allowing us to consider correlation functions for the Virasoro algebra.  A key novelty of the approach described here is that unitary local conformal transformations may be directly simulated and errors estimated. We focus on a fermionic example with central charge $c=\frac12$, however, we also explain how to extend our results to cover currents for Wess-Zumino-Witten models. Our approach admits natural generalisations to a wider class of models with arbitrary central charge, including bosonic models and models based on anyonic systems.  

\paragraph*{Representing conformal fields on a quantum computer.}
We exploit the Hamiltonian formulation of conformal field theory and work with units so that $\hbar = c = 1$ \cite{francescoConformalFieldTheory1997}. In order to simplify the exposition we illustrate our approach using the key example of relativistic Dirac fermions on the circle with circumference $2L$. Depending on the boundary conditions -- i.e., either periodic or antiperiodic -- one realises the \emph{Ramond} (respectively, \emph{Neveu-Schwarz}) sector. The real chiral components of the massless theory give rise to CFTs with central charge $c=\frac12$ \cite{francescoConformalFieldTheory1997}.

To simulate a CFT on a quantum computer we need to approximate the system with a finite number of qubits. This is achieved by discretising the system onto a finite spatial lattice $\Lambda_{N} = \varepsilon_{N}\{-L_{N}, -L_{N} +1, \ldots, L_{N}-2, L_{N}-1\}\subset\varepsilon_{N}  \mathbb{Z}$, where $\varepsilon_NL_{N} = L$. 
Exploiting the staggered lattice approximation we may discretise the theory with the following lattice hamiltonian:
\begin{equation}
\label{eq:stagH}
H^{(N)}_{0} = \varepsilon_{N}^{-1}\tfrac{L}{\pi}\sum_{x\in\Lambda_{N}}\left({\psi^{(1)}_{x+\varepsilon_{N}}}^\dagger{\psi^{(2)}_{x}} - {\psi^{(1)}_{x}}^\dagger{\psi^{(2)}_{x}} + \textup{h.c.} + \lambda_{N}\left({\psi^{(1)}_{x}}^\dagger {\psi^{(1)}_{x}} - {\psi^{(2)}_{x}}^\dagger{\psi^{(2)}_{x}}\right)\right),
\end{equation}
where $\lambda_N$ is the lattice mass coupling constant (the prefactor has been chosen for consistency with the Koo-Saleur formula). Note that $N$ refers to the \emph{spatial scale}, i.e., the negative logarithm of the lattice spacing. The total number $n$ of lattice sites at scale $N$ is given by $n=2^{N+1}$. The lattice fermion operators $\psi^{(j)}_x$, $j=1,2$, obey the standard canonical anticommutation relations: $\{\psi^{(j)}_x, {\psi^{(k)}_y}^\dagger\} = \delta^{j,k}\delta_{x,y}$, with all other anticommutators vanishing. This particular lattice model is exactly solvable via a Fourier transform followed by a Bogoliubov transformation\footnote{Although the lattice model is exactly solvable, the existence of, and simulation of, the continuum limit is still extremely nontrivial, see e.g., \cite{ziniConformalFieldTheories2018} for an alternative approach.}.

The Hilbert space $\mathcal{H}_N$ of states for the lattice is given by fermionic Fock space, with dimension $2^{2n}$. There are now many methods to represent such fermionic systems on a quantum computer, ranging from the original Jordan-Wigner transformation through to the sophisticated methods of Bravyi and Kitaev and Ball-Cirac-Verstraete \cite{jordanUberPaulischeAquivalenzverbot1928,bravyiFermionicQuantumComputation2002,ballFermionsFermionFields2005,verstraeteMappingLocalHamiltonians2005}. For simplicity of exposition we represent fermionic Fock space on a quantum computer using $2n$ qubits, exploiting a standard Jordan-Wigner transformation; we define for $j\in \Lambda_{N+1}$ the operators $a_j \equiv \psi_j^{(1)}$, $a_{j+\varepsilon_{N+1}} \equiv {\psi_j^{(2)}}^\dag$, $a_j \equiv \sigma^x_{-L}\sigma_{-L+\varepsilon_{N+1}}^x\cdots \sigma^x_{j-\varepsilon_{N+1}}\tfrac{1}{2}(\sigma_j^z+i\sigma_j^y)$. Here the fermions are defined with respect to the lattice $\Lambda_{N+1}$ with half the original lattice spacing. In this way the Hamiltonian may be simulated via the following quantum spin system \cite{osborneConformal2021}
\begin{equation}\label{eq:qsham}
	H_0^{(N)} = \varepsilon_{N}^{-1}\tfrac{L}{2\pi}\sum_{j\in \Lambda_{N+1}} \left(\sigma^z_j\sigma^z_{j+\varepsilon_{N+1}} - \sigma^y_j\sigma^y_{j+\varepsilon_{N+1}} + \lambda_{N}\sigma_j^x \right)+ \text{B.C.s}.
\end{equation}
The term ``B.C.s'' refers to the nontrivial term $-\frac12\varepsilon_{N}^{-1}(\pm1 \pm(-1)^F)(\sigma^z_{-L}\sigma^z_{L-\varepsilon_{N+1}} - \sigma^y_{-L}\sigma^y_{L-\varepsilon_{N+1}})$ encoding the possible boundary conditions (with the first sign referring to spins and the second to fermions). It is noteworthy that anti-periodic fermions complemented by periodic spins lead to (1) a vanishing boundary term on states with even fermion parity, specifically the Fock vacuum, and (2) a non-degenerate ground state in the massless case (Neveu-Schwarz sector). We note that there is considerable room for optimisation here, and one can expect that more sophisticated representations will deliver improved resource requirements \cite{cadeStrategiesSolvingFermiHubbard2020}. 

\paragraph*{Description of the quantum simulation algorithm.}
After choosing a desired accuracy $\delta$ for the result of the simulation the algorithm consists of the following four steps, which we describe and analyse in the sequel.
\begin{enumerate}
  \item Prepare the initial state of the lattice. There is a wide class of possibilities, including the ground state, which can be directly achieved using an efficient quantum circuit.  
  \item Optionally apply field operators.
  \item Evolve the system according to a discretised local conformal symmetry transformation, via the Suzuki-Trotter decomposition.  
  \item Exploit phase estimation to measure local field operators, either in real or momentum space.
\end{enumerate}

\paragraph*{Complexity.}
The computational complexity, i.e., the asymptotic scaling of the number of quantum gates required to compute a correlation function, of the algorithm we present here depends on several parameters. The most important is the desired accuracy $\delta$ of the simulation, which measures the absolute difference between the computed discretised approximate results and the theoretical target value
\begin{equation*}
  |\langle \Omega_N|\widetilde{A} \widetilde{B}_t |\Omega_N\rangle - \langle \Omega|A B_t |\Omega\rangle| \le \delta,
\end{equation*}
where $|\Omega\rangle$ is the CFT state (which need not be the vacuum), $A$ and $B$ are field observables, local in either real or momentum space, the subscript $t$ denotes that the observable has been evolved in the Heisenberg picture according to the dynamics generated by a local conformal transformation, e.g., $B_t\equiv e^{it L_k} B e^{-it L_k}$, where $L_k$ is a Virasoro generator, the tildes represent discretised approximations, and $N = \log_{2}(n)-1$ represents the number of lattice sites of the discretised approximation. By increasing the number $n$ of lattice sites one can reduce the errors according to $1/n^{2}$: this provides the direct connection between the complexity of the simulation and the accuracy. There is an implied constant here which depends on the order $k$ of the local conformal transformation desired. One can generalise these error bounds to cover multipoint correlation functions with $m$ operators; a simple bound shows that the errors grow quadratically with $m$. An important feature of our approach is that $A$ and $B$ are \emph{not} required to be bounded operators in the continuum limit: our results also apply to the case where, e.g., $A$ and $B$ are unbounded Virasoro generators. 

It is important to note that there is no mass gap determining a length scale for CFTs. Since our simulation algorithm directly targets the continuum model the length scale for the problem is instead determined by the desired spatial or frequency localization of the field operators appearing in a correlation function. This is manifested in the algorithm as follows: Having chosen an accuracy $\delta$ for the desired correlation function we choose \emph{two} scales, $M < N$ which are large enough to ensure that the discretisation errors incurred are less than $\delta$. The first scale $M$ represents the length scale -- the spatial or momentum \emph{localization} -- of the desired observables and $N$ represents the ultimate \emph{ultraviolet lattice cutoff} of the quantum simulation. Thus we require $2n = 2^{N+2}$ qubits to carry out a simulation of observables defined at a length scale $M$.

\begin{figure}[ht]
\begin{center}
\begin{tikzpicture}
	
	\draw (0.25,0.25) node[right]{$\omega^{(N)}_{0}$} (1.75,0.25) node[right]{$\omega^{(N)}_{1}$} (3.5,0.25) node[right]{$\omega^{(N)}_{2}$}
	(5.5,0.25) node[right]{$\dots$} (6.25,0.25) node[right]{$\omega^{(N)}_\infty$} (7.5,0.25) node[right]{$\mathfrak{A}_N$};
	\draw (1.45,0.25) node[below]{,};
	\draw (3.2,0.25) node[below]{,};
	\draw (4.95,0.25) node[below]{,};
	
	\draw (1.75,1.5) node[right]{$\omega^{(N+1)}_{0}$} (3.5,1.5) node[right]{$\omega^{(N+1)}_{1}$} (5.5,1.5)
	node[right]{$\dots$} (6.25,1.5) node[right]{$\omega^{(N+1)}_\infty$}  (7.5,1.5) node[right]{$\mathfrak{A}_{N+1}$};
	\draw (3.2,1.5) node[below]{,};
	\draw (4.95,1.5) node[below]{,};
	
	\draw (2,0.875) node[right]{$\mathcal{E}^{N+1}_{N}$} (3.75,0.875) node[right]{$\mathcal{E}^{N+1}_{N}$};
	\draw[->] (2,1.2) to (2,0.6);
	\draw[->] (3.75,1.2) to (3.75,0.6);
	\draw[<-] (7.75,1.2) to (7.75,0.6); 
	\draw (7.75,0.875) node[right]{$\alpha^{N}_{N+1}$};
	
	\draw (3.5,2.75) node[right]{$\omega^{(N+2)}_{0}$} (5.5,2.75)
	node[right]{$\dots$} (6.25,2.75) node[right]{$\omega^{(N+2)}_\infty$} (7.5,2.75) node[right]{$\mathfrak{A}_{N+2}$};
	\draw (4.95,2.75) node[below]{,};
	
	\draw (3.75,2.125) node[right]{$\mathcal{E}^{N+2}_{N+1}$};
	\draw[->] (3.75,2.45) to (3.75,1.85);
	\draw[<-] (7.75,2.45) to (7.75,1.85);
	\draw (7.75,2.125) node[right]{$\alpha^{N+1}_{N+2}$};
	
	\node[right] at (4.25,3.475) {$\iddots$};
	\node[right] at (6.375,3.475) {$\vdots$};
	\node[right] at (7.575,3.475) {$\vdots$};
\end{tikzpicture}
\caption{Operator algebraic analogue of Wilson's triangle of renormalization \cite{wilsonRenormalizationGroupCritical1975, stottmeisterOperatoralgebraicRenormalizationWavelets2020}: Vertical lines represent renormalization steps, either by coarse graining states ($\cE$'s) or by refining fields ($\alpha$'s). Horizontal lines represent sequences of renormalized states at a fixed scale.}
\label{fig:statetrianglerg}
\end{center}
\end{figure}
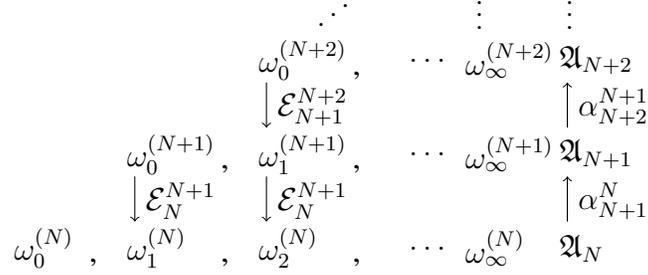

\paragraph*{Discretised representations and renormalization.}
\emph{Operator-algebraic renormalization} \cite{stottmeisterOperatoralgebraicRenormalizationWavelets2020,morinelliScalingLimitsLattice2020,brothierCanonicalQuantization1dimensional2019,osborneContinuumLimitsQuantum2019} is a sophisticated new implementation of Wilsonian renormalization in the Heisenberg picture. We use OAR to build the continuum limit as a sequence of discretised approximations which, in turn, supply us with the lattice models we simulate. The salient aspects of OAR may be summarised as follows. We consider a family, $\{\fA_{N},\cH_{N},H^{(N)}_{0}\}_{N}$, of quantum lattice systems at (spatial) resolution $\vep$ with $N\!=\!-\log_2(\vep)$. At each scale $N$, we have an \emph{algebra} $\fA_{N}$ of observables acting on the Hilbert space $\cH_{N}$ of the lattice, with $H^{(N)}_{0}$ the corresponding Hamiltonian. The renormalization group then provides us with a means to compare the lattice systems at different scales, $N_1 < N_2$, realized with a \emph{coarse-graining operation}, $\cE^{N_{2}}_{N_{1}}\!(\rho^{(N_{2})}_{0})\!=\!\rho^{(N_{1})}_{N_{2}-N_{1}}$, between states $\rho^{(N)}$ at different scales. Here $\mathcal{E}$ is interpreted in the \emph{Schrödinger picture}, however, one may equivalently describe the RG in the \emph{Heisenberg picture}, where the coarse-graining map, now called the \emph{scaling map}, acts dually on operators: $\alpha^{N_{1}}_{N_{2}}:\fA_{N_{1}}\rightarrow\fA_{N_{2}}$. We denote by $\rho^{(N)}_{0}$ the \emph{initial state}, which may be understood as a state for the finest lattice, with lattice spacing $\vep = 2^{-N}$, considered in the sequence, and $\rho^{(N)}_{M}$ is its $M$th renormalization. Demanding equality of the partition functions of the initial and renormalized states requires $\alpha^{N_{1}}_{N_{2}}$ to be unital and completely positive (ucp), i.e.~it preserves probability and sends states to states.
The \emph{renormalization group} is then the collection $\{\alpha^{N_{1}}_{N_{2}}\}_{N_{2}>N_{1}}$, assumed to satisfy $\alpha^{N_{2}}_{N_{3}}\circ\alpha^{N_{1}}_{N_{2}}\!=\!\alpha^{N_{1}}_{N_{3}}$, see Fig.~\ref{fig:statetrianglerg}.
A \textit{scaling limit} $\rho^{(N)}_{\infty}$ is a limit state $\lim_{M\rightarrow\infty}\rho^{(N)}_{M}\!=\!\rho^{(N)}_{\infty}$ on $\fA_N$, which is stable under coarse-graining\footnote{Note that we have described everything in terms of trace-class density operators, however, the considerations here may be extended directly to states on $C^*$ algebras without a density-operator representation. See \cite{osborneConformal2021} for further information.}:
\begin{equation}
\label{eq:staterglimit}
\mathcal{E}_{N_1}^{N_2}(\rho^{(N_{2})}_{\infty}) = \rho^{(N_{1})}_{\infty}.
\end{equation}
Note that finding a limit state usually requires that the coupling constants of the bare model be redefined at each scale according to the \emph{renormalization conditions}.
We are ultimately interested in the \emph{double scaling limit} $\rho_{\infty}^{(\infty)}\equiv\lim_{N\rightarrow \infty} \rho_\infty^{(N)}$, which realises the continuum limit (this exists as soon as the scaling limit exists because of \eqref{eq:staterglimit}, so this does not entail additional constraints on the construction). The scaling limit of the family $H^{(N)}_{0}$ of lattice Hamiltonians is naturally understood via a limit, $\tau^{(\infty)}_{t}$, in the Heisenberg picture of the discretised approximations $\tau^{(N)}_{t}\!=\!e^{i t H^{(N)}_{0}}\!(\!\ \cdot\!\ )e^{-i t H^{(N)}_{0}}$. Thus the triple $\{\fA,\rho,\tau_{t}\}$, where $\fA \equiv \fA_{\infty}$, $\rho \equiv \rho^{(\infty)}_{\infty}$, and $\tau_t \equiv \tau_t^{(\infty)}$, is called a \emph{scaling limit} of $\{\fA_{N},\cH_{N},H^{(N)}_{0}\}_{N}$.

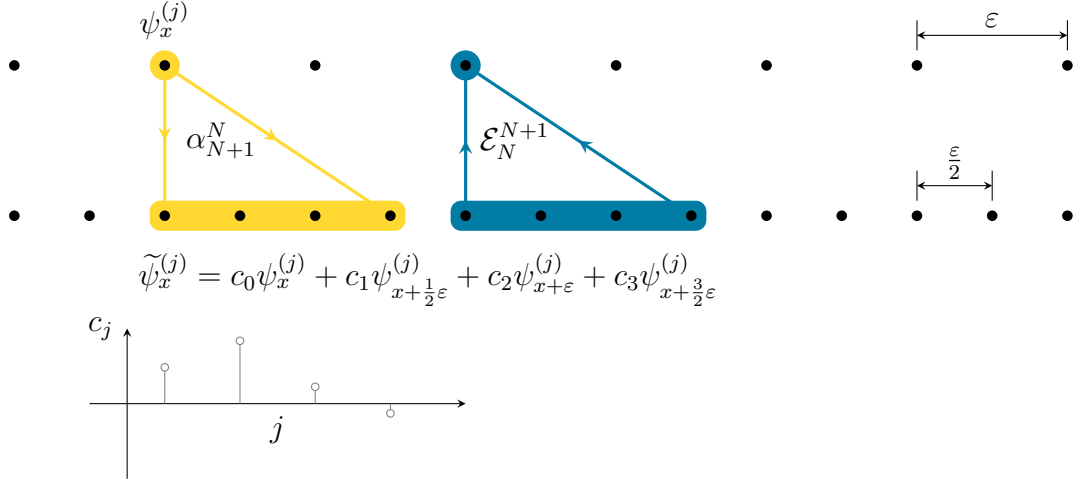
\begin{figure}
	\begin{center}
		\begin{tikzpicture}
			\begin{scope}[very thick,decoration={markings, mark=at position 0.5 with {\arrow{stealth}}}] 
		  	\draw[postaction={decorate}, color=banana] (2,2) -- (2,0);
				\draw[postaction={decorate}, color=banana] (2,2) -- (5,0);
		  	\draw[postaction={decorate}, color=cgblue] (6,0) -- (6,2);
				\draw[postaction={decorate}, color=cgblue] (9,0) -- (6,2);		
			\end{scope}
			\def\x{0.2};
			\def\y{2.5};
			\draw[stealth-stealth] (12,2+2*\x) -- (14,2+2*\x);
			\draw (12,2+\x) -- (12,2+3*\x);
			\draw (14,2+\x) -- (14,2+3*\x);
			\draw (13,2+3*\x) node {$\vep$};
			\draw[stealth-stealth] (12,2*\x) -- (13,2*\x);
			\draw (12,\x) -- (12,3*\x);
			\draw (13,\x) -- (13,3*\x);
			\draw (12.5,3.5*\x) node {$\frac{\vep}{2}$};
			\draw (2.75,1) node {$\alpha^{N}_{N+1}$};
			\draw (6.65,1) node {$\mathcal{E}^{N+1}_{N}$};
			\draw (2,2+3*\x) node {$\psi_x^{(j)}$};
			\draw[rounded corners, draw=none, fill=banana] (2-\x, -\x) rectangle (5+\x, \x) {};
			\draw[draw=none, fill=banana] (2,2) circle (0.2);
			\draw[rounded corners, draw=none, fill=cgblue] (6-\x, -\x) rectangle (9+\x, \x) {};
			\draw[draw=none, fill=cgblue] (6,2) circle (0.2);
			\foreach \j in {0, 2, ..., 15} {
				\fill (\j,2) circle (2pt);
			}
			\foreach \j in {0, 1, ..., 14} {
				\fill (\j,0) circle (2pt);
			}
			\draw[-stealth] (1,-\y) -- (6,-\y); 
			\draw[-stealth] (1.5,-\y-1) -- (1.5,-\y+1); 
			\draw[color=gray] (2,-\y) -- (2,0.4830-\y);
			\draw[color=gray,fill=white] (2,0.4830-\y) circle (0.05);
			\draw[color=gray] (3,-\y) -- (3,0.8365-\y);
			\draw[color=gray,fill=white] (3,0.8365-\y) circle (0.05);
			\draw[color=gray] (4,-\y) -- (4,0.2241-\y);
			\draw[color=gray,fill=white] (4,0.2241-\y) circle (0.05);
			\draw[color=gray] (5,-\y) -- (5,-0.1294-\y);
			\draw[color=gray,fill=white] (5,-0.1294-\y) circle (0.05);
			\draw (3.5,-0.35-\y) node {$j$};
			\draw (1.15,1-\y) node {$c_j$};
			\draw (5.5,-0.85) node {$\widetilde{\psi}_{x}^{(j)}=c_0 \psi_{x}^{(j)} + c_1 \psi_{x+\frac{1}{2}\vep}^{(j)} + c_2 \psi_{x+\vep}^{(j)} + c_3 \psi_{x+\frac{3}{2}\vep}^{(j)}$};
		\end{tikzpicture}
	\caption{Operator algebraic renormalization in real space for the wavelet D$4$. Here $\vep$ labels the lattice spacing of the coarse lattice, $\alpha^{N}_{N+1}$ and $\mathcal{E}^{N+1}_{N}$ denote the RG maps in the Heisenberg (respectively, Schr\"odinger) picture, and $c_j$ are the low-pass filter coefficients defining the D$4$ wavelet.}\label{fig:oar}
	\end{center}
\end{figure}

To exploit OAR we need to select a family of coarse-graining/scaling maps $\mathcal{E}$ which are compatible with the physics we expect in the continuum scaling limit; the success of OAR requires that one retains enough of the quantum coherence present in low-energy degrees of freedom. Two particularly powerful procedures are realised by real-space and momentum-space renormalization. To describe these RG families we note that the algebra $\mathfrak{A}_N$ of observables for the lattice at scale $N=-\log_2(\vep)$ is given by the canonical anticommutation relations (CAR) algebra generated by the lattice fermion operators $\mathfrak{A}_N \equiv \langle \psi_x^{(j)}\,|\, x\in \Lambda_N, j = 1,2\rangle$. We then write the RG maps for our real-space scheme according to 
\begin{equation}\label{eq:waveletrg}
	\alpha_{N+1}^N(\psi_x^{(j)}) = \sum_{l\in \mathbb{Z}} c_l \psi_{x-\vep_{N+1}l}^{(j)},
\end{equation}
where $\{c_l\}_{l\in \mathbb{Z}}$ is the \emph{low-pass filter} determined by a scaling function $s$ associated with (compactly supported) orthogonal wavelets\footnote{Only finitely many of the $c_l$ are nonzero, so that the sum appearing in (\ref{eq:waveletrg}) is well defined as long as $N$  is larger than the number of nonzero terms.} \cite{daubechiesTenLecturesWavelets1992}
\begin{equation}
	s(x) = \sqrt{2} \sum_{l\in\mathbb{Z}}  c_l s(2x -l).
\end{equation}
An explicit example is furnished by the Daubechies D4 wavelet, for which the coefficients are given by $c_0 = \frac{1}{4\sqrt{2}}(1+\sqrt{3})$, $c_1 = \frac{1}{4\sqrt{2}}(3+\sqrt{3})$, $c_2 = \frac{1}{4\sqrt{2}}(3-\sqrt{3})$, and $c_3 = \frac{1}{4\sqrt{2}}(1-\sqrt{3})$. However, in order to simulate observables which, in the continuum limit, are unbounded, it can be necessary to choose wavelets with higher regularity than D4.
Note that the renormalized fermion operator $\widetilde{\psi}_x^{(j)} \equiv \alpha_{N+1}^N(\psi_x^{(j)})$ acts on the fine-grained lattice, with lattice spacing $\frac12\vep$. The resulting real-space OAR is illustrated in Fig.~\ref{fig:oar}.

To describe the momentum-space RG maps we first introduce the Fourier transformed operators
\begin{equation}
	\widehat{\psi}_k^{(j)} \equiv \sum_{x\in\Lambda_N} e^{-ikx}\psi_x^{(j)},
\end{equation}
according to which the RG acts via
\begin{equation}
	\alpha_N^M(\widehat{\psi}_k^{(j)}) = 2^{\frac12(N-M)} \chi_M(k) \widehat{\psi}_k^{(j)},
\end{equation}
where $\chi_M(k) = 1$ if $k\in \frac{\pi}{L}\{-L_{M}+\frac12, \ldots, L_{M}+\frac12\}$ and $0$ otherwise. That is, only  operators $\widehat{\psi}_k^{(j)}$ associated to long wavelengths up to a momentum cutoff $|k| \le \frac{\pi L_{M}}{L}$ are retained, with short wavelength degrees of freedom $|k|>\frac{\pi L_{M}}{L}$ traced out. In both the case of real-space and momentum-space OAR one can check that the RG maps $\alpha_N^M$ satisfy the composition rules $\alpha^{N_{2}}_{N_{3}}\circ\alpha^{N_{1}}_{N_{2}}\!=\!\alpha^{N_{1}}_{N_{3}}$.

\paragraph*{Discretised Virasoro generators.} With the basic kinematical and dynamical data in hand we turn now to the quantum simulation of conformal symmetries. These form the \emph{conformal group} with the associated \emph{Virasoro algebra} of generators spanned by  $L_k$, $k\in\frac{\pi}{L}\mathbb{Z}$, satisfying 
\begin{equation}
	[L_k,L_l] = \tfrac{L}{\pi}(k-l)L_{k+l} + \frac{c}{12}((\tfrac{L}{\pi}k)^3-(\tfrac{L}{\pi}k))\delta_{k+l,0},
\end{equation}
where $c$ is the \emph{central charge}. Several of the generators are directly related to more familiar observables, for example, the CFT Hamiltonian and total momentum are given by
\begin{equation}
	H^{\text{CFT}} = \frac{\pi}{L}\left(L_0+\overline{L}_0 -\frac{c}{12}\right) \quad \text{and}\quad P^{\text{CFT}} = \frac{\pi}{L}\left(L_0-\overline{L}_0\right).
\end{equation}
To approximate $L_k$ on a lattice at scale $N$ we exploit an insightful ansatz due to Koo and Saleur \cite{kooRepresentationsVirasoroAlgebra1994b}; we begin by introducing the fourier modes of the lattice Hamiltonian density\footnote{In the case of massless Dirac fermions ($\lambda_{N}=0$): $h_x^{(N)} = \frac{1}{\vep_N^2}\Big({\psi^{(1)}_{x+\varepsilon_{N}}}^\dagger{\psi^{(2)}_{x}} - {\psi^{(1)}_{x}}^\dagger{\psi^{(2)}_{x}} + {\psi^{(2)}_{x-\varepsilon_{N}}}^\dagger{\psi^{(1)}_{x}} - {\psi^{(2)}_{x}}^\dagger{\psi^{(1)}_{x}} + \textup{h.c.}\Big)$.}, which is symmetric around $x$, to obtain the correct Virasoro algebra in the scaling limit:
\begin{equation}
	H_k^{(N)} \equiv \vep_N \tfrac{L}{\pi} \sum_{x} e^{ikx} h_x^{(N)},
\end{equation} 
and then defining 
\begin{equation}\label{eq:koosaleur}
	L_k^{(N)}  \equiv \frac{1}{2}\left(H_k^{(N)}+ \frac{\pi \vep_N}{2 L \sin(\frac{1}{2}\vep_N k)}[H_k^{(N)},H_0^{(N)}]\right)+\frac{c}{24}\delta_{k,0},
\end{equation}
with an analogous formula for $\overline{L}_k^{(N)}$. The ansatz (\ref{eq:koosaleur}) leads to realisations of CFTs with $c=0$, $c=1/2$, and $c=1$ with the specific value of $c$ being related to the choice of the representation. For $c=0$, we use the defining representation of the CAR algebra corresponding to the ground state associated with a vanishing stress-energy tensor. Central charges $c=1$ and $c=1/2$ arise from the ground state of the massless free fermion Hamiltonian, the first from a representation of a complex chiral component, the second from a representation of a real (or self-dual) chiral component via the projection on positive (or negative) momenta \cite{osborneConformal2021}.

The crucial observation underlying our quantum simulation algorithm is that $L_k^{(N)}$ so defined in (\ref{eq:koosaleur}) are sums of densities. This means that we may employ standard quantum algorithms for the simulation of the dynamics of local hamiltonians to simulate the exponentials of these generators\footnote{In order to ensure that the resulting dynamics are unitary we actually simulate hermitian linear combinations of $L_k^{(N)}$ and $\overline{L}_k^{(N)}$.}; since the behaviour of these methods are well understood we focus on the new challenges arising in approximating the continuum limit with a lattice model.

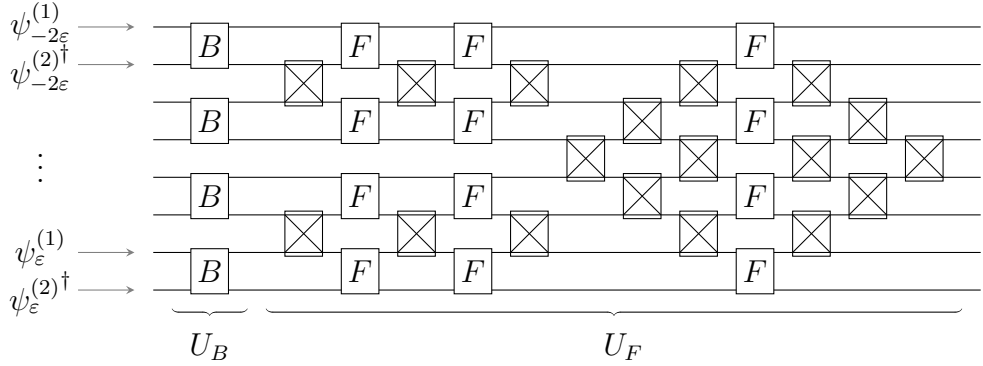
\begin{figure}
	\begin{center}
		\begin{tikzpicture}
			\def\x{0.05};
			\def\y{2.5};
			\draw (-2,3.5+\x) node {$\psi^{(1)}_{-2\vep}$};
			\draw[color=gray,-stealth] (-1.5, 3.5) -- (-0.75, 3.5); 
			\draw (-2,3-\x) node {${\psi^{(2)}_{-2\vep}}^{\!\!\!\dag}$};
			\draw[color=gray,-stealth] (-1.5, 3) -- (-0.75, 3); 
			\draw (-2,0.5+\x) node {$\psi^{(1)}_{\vep}$};
			\draw[color=gray,-stealth] (-1.5, 0.5) -- (-0.75, 0.5); 
			\draw (-2,0-\x) node {${\psi^{(2)}_{\vep}}^\dag$};
			\draw[color=gray,-stealth] (-1.5, 0) -- (-0.75, 0); 
			\draw (-2,1.75) node {$\vdots$};
			\foreach \j in {0, 0.5, ..., 3.5} {
				\draw (-0.5,\j) -- (10.5,\j);
			}
			\foreach \j in {0, 1, ..., 3.5} {
				\draw[fill = white] (0,\j-\x) rectangle (0.5, \j+0.5+\x);
				\draw (0.25, \j+0.25) node {$B$};
			}
			\foreach \j in {0.5,2.5} {
				\draw[fill = white] (1.25,\j-\x) rectangle (1.75, \j+0.5+\x);
				\draw (1.25,\j) -- (1.75,\j);
				\draw (1.25,\j+0.5) -- (1.75,\j+0.5);
				\draw (1.25,\j) -- (1.75,\j+0.5);
				\draw (1.25,\j+0.5) -- (1.75,\j);				
			}
			\foreach \j in {0, 1, ..., 3.5} {
				\draw[fill = white] (2,\j-\x) rectangle (2.5, \j+0.5+\x);
				\draw (2.25, \j+0.25) node {$F$};
			}
			\foreach \j in {0.5,2.5} {
				\draw[fill = white] (2.75,\j-\x) rectangle (3.25, \j+0.5+\x);
				\draw (2.75,\j) -- (3.25,\j);
				\draw (2.75,\j+0.5) -- (3.25,\j+0.5);
				\draw (2.75,\j) -- (3.25,\j+0.5);
				\draw (2.75,\j+0.5) -- (3.25,\j);				
			}
			\foreach \j in {0, 1, ..., 3.5} {
				\draw[fill = white] (3.5,\j-\x) rectangle (4, \j+0.5+\x);
				\draw (3.75, \j+0.25) node {$F$};
			}
			\foreach \j in {0.5,2.5} {
				\draw[fill = white] (4.25,\j-\x) rectangle (4.75, \j+0.5+\x);
				\draw (4.25,\j) -- (4.75,\j);
				\draw (4.25,\j+0.5) -- (4.75,\j+0.5);
				\draw (4.25,\j) -- (4.75,\j+0.5);
				\draw (4.25,\j+0.5) -- (4.75,\j);				
			}
			\foreach \j in {1.5} {
				\draw[fill = white] (5,\j-\x) rectangle (5.5, \j+0.5+\x);
				\draw (5,\j) -- (5.5,\j);
				\draw (5,\j+0.5) -- (5.5,\j+0.5);
				\draw (5,\j) -- (5.5,\j+0.5);
				\draw (5,\j+0.5) -- (5.5,\j);	
			}
			\foreach \j in {1, 2} {
				\draw[fill = white] (5.75,\j-\x) rectangle (6.25, \j+0.5+\x);
				\draw (5.75,\j) -- (6.25,\j);
				\draw (5.75,\j+0.5) -- (6.25,\j+0.5);
				\draw (5.75,\j) -- (6.25,\j+0.5);
				\draw (5.75,\j+0.5) -- (6.25,\j);	
			}
			\foreach \j in {0.5, 1.5, 2.5} {
				\draw[fill = white] (6.5,\j-\x) rectangle (7, \j+0.5+\x);
				\draw (6.5,\j) -- (7,\j);
				\draw (6.5,\j+0.5) -- (7,\j+0.5);
				\draw (6.5,\j) -- (7,\j+0.5);
				\draw (6.5,\j+0.5) -- (7,\j);	
			}
			\foreach \j in {0, 1, ..., 3.5} {
				\draw[fill = white] (7.25,\j-\x) rectangle (7.75, \j+0.5+\x);
				\draw (7.5, \j+0.25) node {$F$};
			}
			\foreach \j in {0.5, 1.5, 2.5} {
				\draw[fill = white] (8,\j-\x) rectangle (8.5, \j+0.5+\x);
				\draw (8,\j) -- (8.5,\j);
				\draw (8,\j+0.5) -- (8.5,\j+0.5);
				\draw (8,\j) -- (8.5,\j+0.5);
				\draw (8,\j+0.5) -- (8.5,\j);	
			}
			\foreach \j in {1, 2} {
				\draw[fill = white] (8.75,\j-\x) rectangle (9.25, \j+0.5+\x);
				\draw (8.75,\j) -- (9.25,\j);
				\draw (8.75,\j+0.5) -- (9.25,\j+0.5);
				\draw (8.75,\j) -- (9.25,\j+0.5);
				\draw (8.75,\j+0.5) -- (9.25,\j);	
			}
			\foreach \j in {1.5} {
				\draw[fill = white] (9.5,\j-\x) rectangle (10, \j+0.5+\x);
				\draw (9.5,\j) -- (10,\j);
				\draw (9.5,\j+0.5) -- (10,\j+0.5);
				\draw (9.5,\j) -- (10,\j+0.5);
				\draw (9.5,\j+0.5) -- (10,\j);	
			}
			\draw [decorate, decoration = {calligraphic brace}] (0.75,-0.25) --  (-0.25,-0.25);
			\draw (0.25,-0.75) node {$U_B$};
			\draw [decorate, decoration = {calligraphic brace}] (10.25,-0.25) --  (1,-0.25);
			\draw (5.75,-0.75) node {$U_F$};
		\end{tikzpicture}
	\caption{Illustrating the ground-state preparation circuit for $4$ sites. Depicted on the left are the original fermion operators which are identified with the corresponding sites in the spin model via the Jordan-Wigner mapping.}\label{fig:fft}
	\end{center}
\end{figure}

\paragraph*{Preparing the initial state.} The first step is to initialise the $2n$ qubits of the quantum computer into a discrete approximation of the desired state. We focus, for concreteness, on the approximation $|\Omega_N\rangle$ of the ground state $|\Omega\rangle$ of the scaling limit, and states in the conformal block found by applying Virasoro generators to $|\Omega\rangle$. (It is important to note that our algorithm is by no means limited to such states, and can be applied to any quantum state with a scaling continuum limit.) There are now several approaches to the efficient preparation of $|\Omega_N\rangle$. A particularly parsimonious quantum circuit, which we employ here, is adapted from \cite{verstraeteQuantumCircuitsStrongly2009} (see Fig.~\ref{fig:fft}): Note that the quantum spin Hamiltonian (\ref{eq:qsham}) may be diagonalized by a quantum circuit for a (fast) fourier transform $U_{\text{FT}}$ followed by a Bogoliubov transformation $U_{\text{B}}$, so $H_0^{(N)} = U_{\text{FT}}U_{\text{B}} D^{(N)}_0 U_{\text{B}}^\dag U_{\text{FT}}^\dag$, where $D^{(N)}_0$ is diagonal. In this way we may realise the ground state from a product state:
\begin{equation}
	|\Omega_N\rangle = U_{\text{FT}}U_{\text{B}} \bigotimes_{k\in \Gamma_{N+1}} |\!\leftarrow\rangle.
\end{equation}
With the qubit registers in the discrete approximation of the ground state we now apply the desired generators $A$ and $B$ for the correlation function to state $|\Omega_N\rangle$. For concreteness we focus on the momentum-space RG.
In the case of correlation functions for operators $\widehat{\psi}_k^{(j)}$, with $|k|\le \frac{\pi L_{M}}{L}$, local in momentum space we have that
\begin{equation}
	\widehat{\psi}_k^{(j)} = U_{\text{FT}}\sigma^x_{-\frac{\pi}{L}(-L_{N+1}+\frac12)}\cdots \sigma^x_{k-\frac{\pi}{L}}(\tfrac{1}{2}\sigma_{k}^z-\tfrac{1}{2}(-1)^ji\sigma_{k}^y)U_{\text{FT}}^\dag,
\end{equation}
The action of this nonunitary operator may be simulated, e.g., by introducing an ancilla qubit initialised in the state $\frac{1}{\sqrt{2}}(|0\rangle+|1\rangle\!)$, applying the unitary gate $V\!=\!U_{\text{FT}}\sigma^x_{-\frac{\pi}{L}(-L_{N+1}+\frac12)}\!\cdots \sigma^x_{k-\frac{\pi}{L}}(\sigma_{k}^z\otimes |0\rangle\langle
0|-(-1)^ji\sigma_{k}^y\otimes |1\rangle\langle
1|)U_{\text{FT}}^\dag$ to the momentum-space representation of the ground state $U_{\text{FT}}U_{\text{B}}\bigotimes_{k\in \Gamma_{N+1}} |\!\!\leftarrow\rangle$, and then measuring the $\sigma^x$ operator on the ancilla qubit, postselecting on the positive outcome. Repeating this procedure a constant number of times will ensure the correct state is produced with high probability (for a constant number of field operators). In this way we may easily prepare
the initial state $\prod_{k_l}\widehat{\psi}_{k_l}^{(j_l)}|\Omega_N\rangle$. (The application of a field operator local in real space is more involved and requires consideration of the Wavelet basis employed in the OAR. See \cite{osborneConformal2021} for further details.) The resulting initial state is denoted $|\Phi_{\text{init}}\rangle$.

\paragraph*{Simulating dynamics.} Now that we have prepared the initial state we may apply local conformal transformations. Formulating the Fourier-transformed fermion operators in terms of the single-component fermion, $\psi^{(1)}_{k}=\tfrac{1}{2}(\hat{a}_{k}+\hat{a}_{k+\frac{\pi}{\varepsilon_{N+1}}})$, $\psi^{(2)}_{k}=\tfrac{1}{2}e^{i\varepsilon_{N+1}k}(\hat{a}^{\dag}_{-k}-\hat{a}^{\dag}_{-k+\frac{\pi}{\varepsilon_{N+1}}})$, these are generated by hermitian linear combinations of the following Koo-Saleur operators
\begin{align}
	L_k^{(N)} & = \frac{e^{-\frac{i}{4}\varepsilon_{N}k}}{8\pi}\sum_{l,l'\in\Gamma_{N+1}} \!\!\! \begin{pmatrix}
		\hat{a}_{l'} \\ \hat{a}^{\dag}_{-l'}
	\end{pmatrix}^{\dag}
	\delta_{k,l'-l\!\!\!\!\mod\frac{2\pi}{\varepsilon_{N}}}
	\ell^{(N)}_{k}(l',l) \begin{pmatrix}
		\hat{a}_{l} \\ \hat{a}^{\dag}_{-l}
	\end{pmatrix}, \\ \nonumber
	\ell^{(N)}_{k}(l',l) & = \tfrac{1}{2}\!\begin{pmatrix} \hspace{-3cm} -e^{\frac{i}{4}\varepsilon_{N}k}\sin(\varepsilon_{N}(l+\tfrac{k}{2})\!) & \hspace{-3cm} -i(e^{\frac{i}{4}\varepsilon_{N}k}\sin(\varepsilon_{N}\tfrac{l}{2})+e^{-\frac{i}{4}\varepsilon_{N}k}\sin(\varepsilon_{N}\tfrac{l+k}{2})\!) \\ i(e^{\frac{i}{4}\varepsilon_{N}k}\sin(\varepsilon_{N}\tfrac{l'-k}{2})+e^{-\frac{i}{4}\varepsilon_{N}k}\sin(\varepsilon_{N}\tfrac{l'}{2})) & -e^{-\frac{i}{4}\varepsilon_{N}k}\sin(\varepsilon_{N}\tfrac{l+l'}{2}) \end{pmatrix}
\end{align}
These generators are $(2k+1)$-local in the momentum representation and may be simulated via a Suzuki-Trotter decomposition. (A state-of-the-art implementation will make use of the optimal simulation algorithms.)
After application of the generator the quantum register is in the state
\begin{equation}
	|\Phi'\rangle = e^{-i s (L_k^{(N)}+\overline{L}_k^{(N)})}|\Phi_{\text{init}}\rangle.
\end{equation}
Note that the $k=0$ case reduces to simulating the unitary Hamiltonian evolution.

\paragraph*{Measuring observables.} To measure a desired observable $\widehat{O}$, e.g., $\widehat{O} = \widehat{\psi}_k^{(j)}$, one may employ standard phase estimation \cite{nielsenQuantumComputationQuantum2000,childsQuantumSearchMeasurement2002}. This requires the addition of $r$ ancilla qubits to represent the readout, where $r$ is the required to be large enough to resolve all of the eigenvalues of $\widehat{O}$. Repeating this procedure delivers an estimate for the expectation value
\begin{equation}
	\langle\widehat{O} \rangle = \langle\Phi'|\widehat{O}|\Phi'\rangle.
\end{equation}

\paragraph*{Error analysis.} The errors in the quantum simulation arise from several sources. The first, and easiest to control, is the quantum projection noise arising the estimation of $\langle \widehat{O}\rangle$. This may be reduced in the standard way via parallel repetition. The second salient source of error arises from the quantum simulation of the unitary $e^{-i s (L_k^{(N)}+\overline{L}_k^{(N)})}$ via, e.g., product formulas \cite{childsNearlyOptimalLattice2019}, and may be reduced by exploiting higher-order approximations. Exploiting state-of-the art estimates, one requires quantum gates $(nt)^{1+o(1)}$ to reduce this error to below $\delta/2$. The state preparation is considered exact (up to the gate compilation required to implement the single- and two-qubit unitaries required). The fourth source of error arises from the discretization itself. This is deeply nontrivial and requires an extensive and challenging analysis, discussed in the companion paper \cite{osborneConformal2021}. The result of this analysis implies the existence, for a given desired accuracy $\delta$ and $t\leq T$, of a constant size $N(\delta,T)$ lattice achieving this accuracy. The dependence of $N$ on $\delta$ and $T$ is nontrivial, however, numerical experiments --- see the Supplementary Material --- indicate that, e.g., a D$10$ wavelet can achieve an accuracy of $\delta = 1/2$ on a lattice of 128 sites. (This is for the case where one computes dynamical correlation functions involving Virasoro generators.)

\paragraph*{Extensions.} One may generalise the presented algorithm to approximate the lattice currents of Wess-Zumino-Witten models. For example, to target the level-$1$ WZW model for $\widehat{\mathfrak{u}}(D)_{1}$, we require $D$-component fermion fields, which we can simulate using $D$ copies of the realisation exploited above. Details of this approach may be found in \cite{osborneConformal2021}. The Ising CFT may be directly targeted using a modification of the approach here. The details of this will be presented in a follow up publication. More generally, to simulate an arbtirary rational CFT one must take recourse to models involving chains of anyons. To employ OAR we need an appropriate family of coarse-graining maps. Such a family is induced by the Jones-Wenzl projection, however, this is beyond the scope of this paper and will be presented elsewhere.

\paragraph*{Conclusions and future directions.} We have, in this paper, demonstrated that quantum computers may be exploited to simulate the local conformal dynamics of CFTs. The simulation algorithms are built exploiting operator-algebraic renormalization realising the continuum limit as a sequence of lattice models. Local conformal transformations are simulated via the Koo-Saleur generators. Generalizations to the Ising CFT and WZW models are directly possible.

\bibliography{scibib}

\begin{thebibliography}{10}

\bibitem{douglasFoundationsQuantumField2012}
M.~R. Douglas, {\it Proceedings of {{Symposia}} in {{Pure Mathematics}}\/}
  ({American Mathematical Society}, {University of Pennsylvania, Philadelphia,
  PA}, 2012), vol.~85, pp. 105--124.

\bibitem{seibergNathanSeiberg20152014}
N.~Seiberg, 2015 {{Breakthrough Prize}} in {{Fundamental Physics Symposium}} -
  {{YouTube}}, https://www.youtube.com/watch?v=Hi3e0HVxlFo (2014).

\bibitem{wittenAPSMedalExceptional2018}
E.~Witten, {\it Rev. Modern Phys.\/} {\bf 90}, 045003 (2018).

\bibitem{feynmanSimulatingPhysicsComputers1981}
R.~P. Feynman, {\it Int. J. Theor. Phys.\/} {\bf 21}, 467 (1981).

\bibitem{lloydUniversalQuantumSimulators1996}
S.~Lloyd, {\it Science\/} {\bf 273}, 1073 (1996).

\bibitem{abramsSimulationManyBodyFermi1997a}
D.~S. Abrams, S.~Lloyd, {\it Phys. Rev. Lett.\/} {\bf 79}, 2586 (1997).

\bibitem{zalkaSimulatingQuantumSystems1998}
C.~Zalka, {\it Proc. Roy. Soc. of London Ser. A\/} {\bf 454}, 313 (1998).

\bibitem{smithSimulatingQuantumManybody2019}
A.~Smith, M.~S. Kim, F.~Pollmann, J.~Knolle, {\it npj Quantum Information\/}
  {\bf 5}, 1 (2019).

\bibitem{zhangObservationManybodyDynamical2017}
J.~Zhang, {\it et~al.\/}, {\it Nature\/} {\bf 551}, 601 (2017).

\bibitem{bernienProbingManybodyDynamics2017}
H.~Bernien, {\it et~al.\/}, {\it Nature\/} {\bf 551}, 579 (2017).

\bibitem{jordanQuantumAlgorithmsQuantum2012}
S.~P. Jordan, K.~S.~M. Lee, J.~Preskill, {\it Science\/} {\bf 336}, 1130
  (2012).

\bibitem{hamedmoosavianFasterQuantumAlgorithm2018}
A.~Hamed~Moosavian, S.~Jordan, {\it Phys. Rev. A\/} {\bf 98}, 012332 (2018).

\bibitem{jordanQuantumAlgorithmsFermionic2014}
S.~P. Jordan, K.~S.~M. Lee, J.~Preskill, Quantum {{Algorithms}} for {{Fermionic
  Quantum Field Theories}} (2014). ArXiv:1404.7115.

\bibitem{preskillSimulatingQuantumField2018}
J.~Preskill, {\it The 36th Annual International Symposium on Lattice Field
  Theory - LATTICE2018\/} (2018). ArXiv:1811.10085.

\bibitem{wilsonRenormalizationGroupCritical1975}
K.~G. Wilson, {\it Rev. Modern Phys.\/} {\bf 47}, 773 (1975).

\bibitem{wilsonRenormalizationGroupCritical1983}
K.~G. Wilson, {\it Rev. Modern Phys.\/} {\bf 55}, 583 (1983).

\bibitem{wilsonRenormalizationGroupEpsilon1974}
K.~G. Wilson, J.~B. Kogut, {\it Phys. Rept.\/} {\bf 12}, 75 (1974).

\bibitem{durrInitioDeterminationLight2008}
S.~D{\"u}rr, {\it et~al.\/}, {\it Science\/} {\bf 322}, 1224 (2008).

\bibitem{creutzQuarksGluonsLattices1985}
M.~Creutz, {\it Quarks, Gluons and Lattices\/} ({Cambridge University Press},
  {Cambridge}, 1985).

\bibitem{belavinInfiniteConformalSymmetry1984}
A.~A. Belavin, A.~M. Polyakov, A.~B. Zamolodchikov, {\it Nucl. Phys. B\/} {\bf
  241}, 333 (1984).

\bibitem{francescoConformalFieldTheory1997}
P.~Francesco, P.~Mathieu, D.~S{\'e}n{\'e}chal, {\it Conformal {{Field
  Theory}}\/} ({Springer-Verlag}, 1997), first edn.

\bibitem{ziniConformalFieldTheories2018}
M.~S. Zini, Z.~Wang, {\it Commun. Math. Phys.\/} {\bf 363}, 877 (2018).
  ArXiv:1706.08497.

\bibitem{fredenhagenSuperselectionSectorsBraid1992}
K.~Fredenhagen, K.-H. Rehren, B.~Schroer, {\it Rev. Math. Phys.\/} {\bf 04},
  113 (1992).

\bibitem{borcherdsVertexAlgebrasKacMoody1986}
R.~E. Borcherds, {\it Proc. Nat. Acad. Sci. U.S.A.\/} {\bf 83}, 3068 (1986).

\bibitem{lepowskyIntroductionVertexOperator2004}
J.~Lepowsky, H.~Li, {\it Introduction to {{Vertex Operator Algebras}} and
  {{Their Representations}}\/}, Progress in {{Mathematics}} ({Birkh\"auser
  Basel}, 2004).

\bibitem{segalDefinitionConformalField2004}
G.~Segal, {\it Topology, Geometry and Quantum Field Theory\/} ({Cambridge Univ.
  Press, Cambridge}, 2004), vol. 308 of {\it London {{Math}}. {{Soc}}.
  {{Lecture Note Ser}}.\/}, pp. 421--577.

\bibitem{stottmeisterOperatoralgebraicRenormalizationWavelets2020}
A.~Stottmeister, V.~Morinelli, G.~Morsella, Y.~Tanimoto, Operator-algebraic
  renormalization and wavelets (2020). ArXiv:2002.01442.

\bibitem{honglerConformalFieldTheory2019}
C.~Hongler, F.~J. Viklund, K.~Kyt{\"o}l{\"a}, Conformal {{Field Theory}} at the
  {{Lattice Level}}: {{Discrete Complex Analysis}} and {{Virasoro Structure}}
  (2019). ArXiv:1307.4104.

\bibitem{cardyScalingRenormalizationStatistical1996}
J.~Cardy, {\it Scaling and Renormalization in Statistical Physics\/}
  ({Cambridge University Press}, {Cambridge}, 1996).

\bibitem{maldacenaLargeLimitSuperconformal1998}
J.~Maldacena, {\it Adv. Theor. Math. Phys.\/} {\bf 2}, 231 (1998).

\bibitem{osborneContinuumLimitsQuantum2019}
T.~J. Osborne, Continuum {{Limits}} of {{Quantum Lattice Systems}} (2019).
  ArXiv:1901.06124.

\bibitem{morinelliScalingLimitsLattice2020}
V.~Morinelli, G.~Morsella, A.~Stottmeister, Y.~Tanimoto, Scaling limits of
  lattice quantum fields by wavelets (2020). ArXiv:2010.11121.

\bibitem{baoQuantumAlgorithmsConformal2019}
N.~Bao, J.~Liu, {\it Nucl. Phys. B\/} {\bf 946}, 114702 (2019).
  ArXiv:1811.05675.

\bibitem{daubechiesTenLecturesWavelets1992}
I.~Daubechies, {\it Ten {{Lectures}} on {{Wavelets}}\/} ({Society for
  Industrial and Applied Mathematics}, 1992).

\bibitem{kooRepresentationsVirasoroAlgebra1994b}
W.~M. Koo, H.~Saleur, {\it Nucl. Phys. B\/} {\bf 426}, 459 (1994).

\bibitem{milstedExtractionConformalData2017a}
A.~Milsted, G.~Vidal, {\it Phys. Rev. B\/} {\bf 96}, 245105 (2017).

\bibitem{zouConformalFieldsOperator2020}
Y.~Zou, A.~Milsted, G.~Vidal, {\it Phys. Rev. Lett.\/} {\bf 124}, 040604
  (2020).

\bibitem{zouEmergenceConformalSymmetry2020}
Y.~Zou, G.~Vidal, {\it Phys. Rev. B\/} {\bf 101}, 045132 (2020).

\bibitem{zouConformalDataRenormalization2018a}
Y.~Zou, A.~Milsted, G.~Vidal, {\it Phys. Rev. Lett.\/} {\bf 121}, 230402
  (2018).

\bibitem{konigMatrixProductApproximations2017}
R.~K{\"o}nig, V.~B. Scholz, {\it Nucl. Phys. B\/} {\bf 920}, 32 (2017).

\bibitem{konigMatrixProductApproximations2016}
R.~K{\"o}nig, V.~B. Scholz, {\it Phys. Rev. Lett.\/} {\bf 117}, 121601 (2016).

\bibitem{haegemanRigorousFreeFermionEntanglement2018}
J.~Haegeman, {\it et~al.\/}, {\it Phys. Rev. X\/} {\bf 8}, 011003 (2018).

\bibitem{singhHolographicConstructionQuantum2016}
S.~Singh, G.~K. Brennen, Holographic {{Construction}} of {{Quantum Field
  Theory}} using {{Wavelets}} (2016). ArXiv:1606.05068.

\bibitem{jordanUberPaulischeAquivalenzverbot1928}
P.~Jordan, E.~Wigner, {\it Z. Phys.\/} {\bf 47}, 631 (1928).

\bibitem{bravyiFermionicQuantumComputation2002}
S.~B. Bravyi, A.~Y. Kitaev, {\it Ann. Phys.\/} {\bf 298}, 210 (2002).

\bibitem{ballFermionsFermionFields2005}
R.~C. Ball, {\it Phys. Rev. Lett.\/} {\bf 95}, 176407 (2005).

\bibitem{verstraeteMappingLocalHamiltonians2005}
F.~Verstraete, J.~I. Cirac, {\it J. Stat. Mech.-Theory E.\/} {\bf 2005}, P09012
  (2005).

\bibitem{osborneConformal2021}
T.~J. Osborne, A.~Stottmeister, Conformal field theory from lattice fermions
  (2021). ArXiv:2107.13834.

\bibitem{cadeStrategiesSolvingFermiHubbard2020}
C.~Cade, L.~Mineh, A.~Montanaro, S.~Stanisic, {\it Phys. Rev. B\/} {\bf 102},
  235122 (2020).

\bibitem{brothierCanonicalQuantization1dimensional2019}
A.~Brothier, A.~Stottmeister, Canonical quantization of 1+1-dimensional
  {{Yang}}-{{Mills}} theory: {{An}} operator-algebraic approach (2019).
  ArXiv:1907.05549.

\bibitem{verstraeteQuantumCircuitsStrongly2009}
F.~Verstraete, J.~I. Cirac, J.~I. Latorre, {\it Phys. Rev. A\/} {\bf 79},
  032316 (2009).

\bibitem{nielsenQuantumComputationQuantum2000}
M.~A. Nielsen, I.~L. Chuang, {\it Quantum Computation and Quantum
  Information\/} ({Cambridge University Press}, {Cambridge}, 2000).

\bibitem{childsQuantumSearchMeasurement2002}
A.~M. Childs, {\it et~al.\/}, {\it Phys. Rev. A\/} {\bf 66}, 032314 (2002).

\bibitem{childsNearlyOptimalLattice2019}
A.~M. Childs, Y.~Su, {\it Phys. Rev. Lett.\/} {\bf 123}, 050503.

\end{thebibliography}

\bibliographystyle{Science}

\section*{Acknowledgments}
The authors would like to thank R. F. Werner for valuable discussion about inductive limits
in quantum theory. AS would like to thank Y. Tanimoto for helpful discussions concerning
the essential self-adjointness of smeared Virasoro generators. This work was supported, in part, by the DFG through SFB 1227 (DQ-mat), Quantum Valley Lower Saxony, and funded by the Deutsche Forschungsgemeinschaft (DFG, German Research Foundation) under Germanys Excellence Strategy EXC-2123 QuantumFrontiers 390837967.  

\newpage
\appendix
\section*{Supplementary material}
\newpage
\begin{figure}
	\begin{center}
	\begin{tikzpicture}
	\draw (0,0) node{\includegraphics[width=0.9\textwidth]{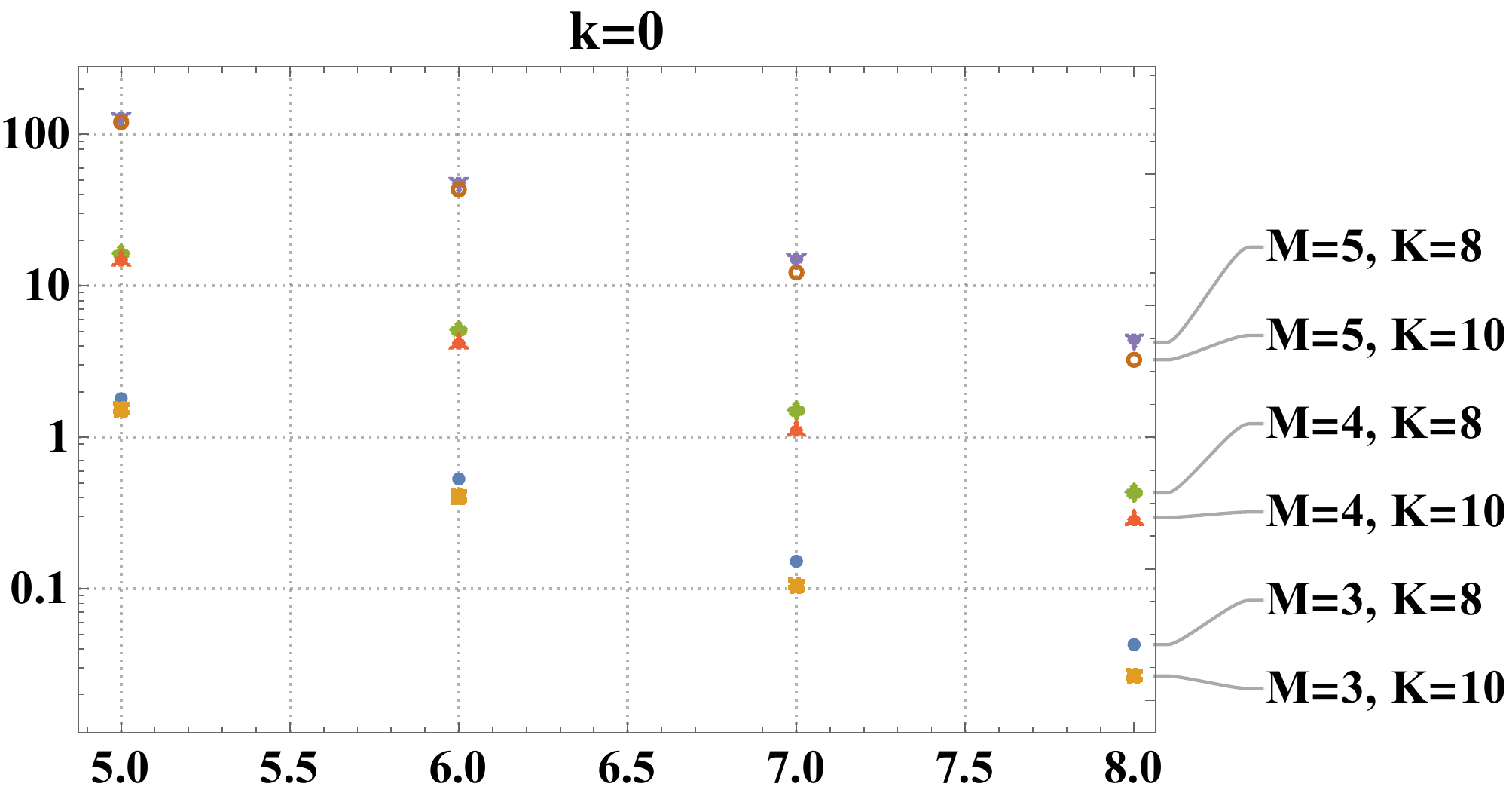}};
	\draw (4.25,-3.15) node{\textbf{N}};
	\draw (-6.35,3.6) node{$\mathbf{L}^{2}$-\textbf{error}};
	\end{tikzpicture}
	\caption{Upper bound to the $L^{2}$-norm error between the conformal Hamiltonian $L_{0}$ and its KS approximant $L^{(N)}_{0}$ at $c=0$ on one-particle states corresponding to the subspace of the CFT Hilbert space representing wavelet details (for Daubechies wavelets D$K$) up to the simulation scale $M$ as a function of the UV cutoff scale $N$. For further details, see \cite{osborneConformal2021}.}\label{fig:waveleterrorM}
	\end{center}
\end{figure}

\begin{figure}
	\begin{center}
	\begin{tikzpicture}
	\draw (0,0) node{\includegraphics[width=0.9\textwidth]{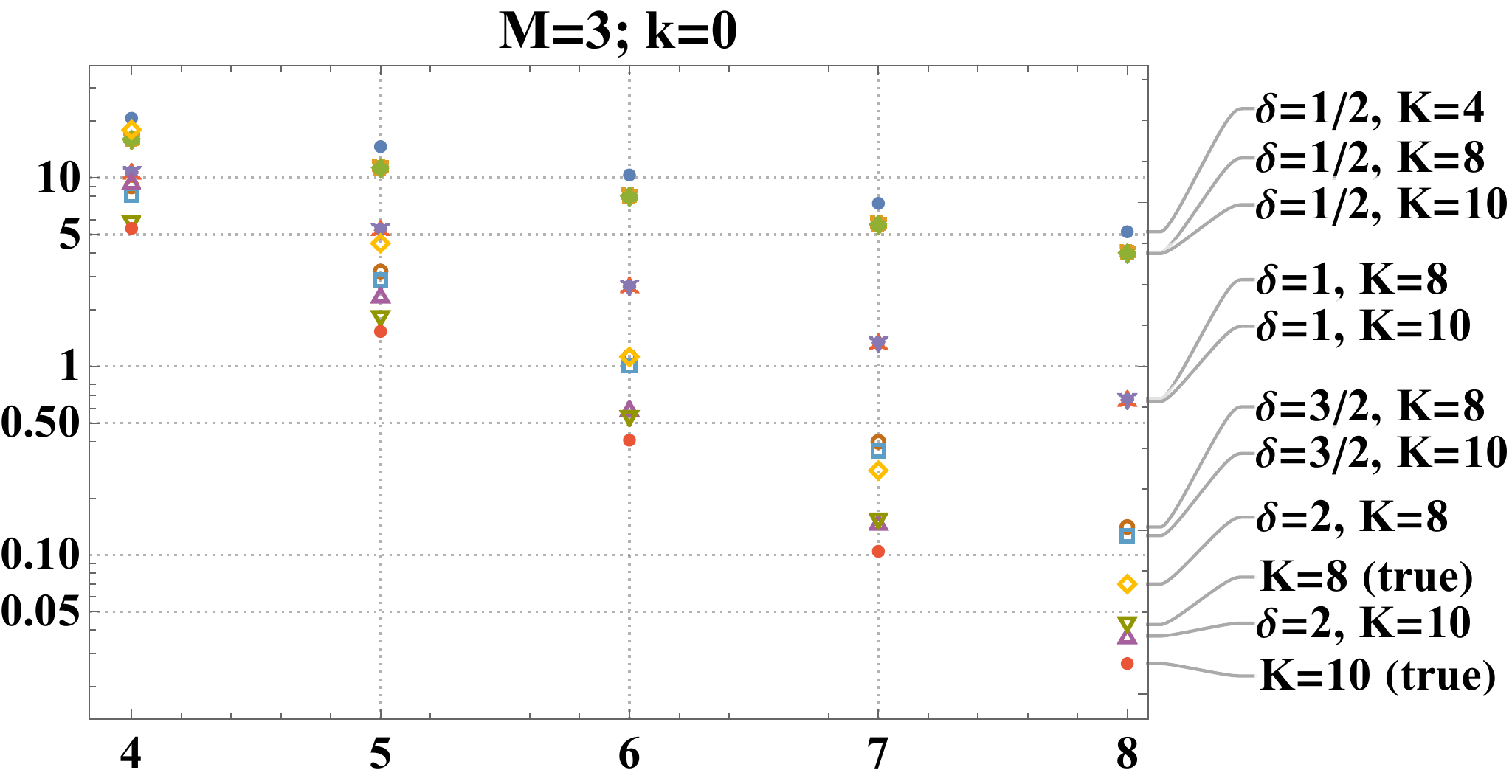}};
	\draw (4.15,-3.1) node{\textbf{N}};
	\draw (-6.3,3.55) node{$\mathbf{L}^{2}$-\textbf{error}};
	\end{tikzpicture}
	\caption{Various upper bounds to the $L^{2}$-norm error between the conformal Hamiltonian $L_{0}$ and its KS approximant $L^{(N)}_{0}$ at $c=0$ on one-particle states corresponding to the subspace of the CFT Hilbert space representing wavelet details (for Daubechies wavelets D$K$) up to the simulation scale $M$ as a function of the UV cutoff scale $N$. The parameter $\delta$ is related to a factorized upper bound containing an $H^{\delta}$-Sobolev norm of the Daubechies D$K$ scaling function. For further details, see \cite{osborneConformal2021}.}\label{fig:waveleterrork}
	\end{center}
\end{figure}

\begin{figure}
	\begin{center}
	\begin{tikzpicture}
	\draw (0,0) node{\includegraphics[width=0.9\textwidth]{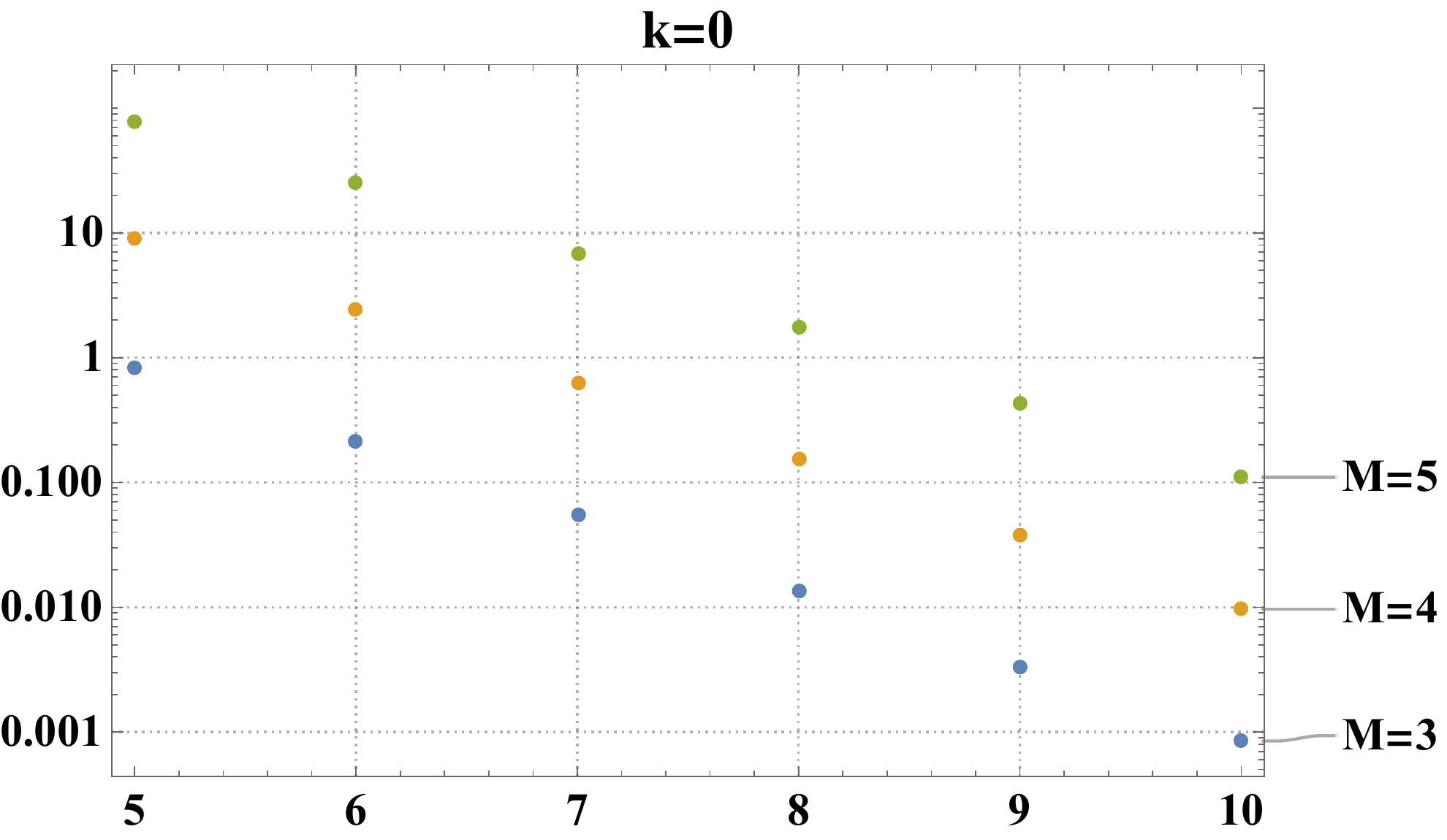}};
	\draw (5.75,-3.55) node{\textbf{N}};
	\draw (-6,4) node{$\mathbf{L}^{2}$-\textbf{error}};
	\end{tikzpicture}
	\caption{Upper bound to the (diagonal) $L^{2}$-norm error between the conformal Hamiltonian $L_{0}$ and its KS approximant $L^{(N)}_{0}$ at $c=1/2,1$ on one-particle states corresponding to the subspace of the CFT Hilbert space representing momenta up to the simulation scale $M$ as a function of the UV cutoff scale $N$. For further details, see \cite{osborneConformal2021}.}\label{fig:momrgerrorM}
	\end{center}
\end{figure}

\begin{figure}
	\begin{center}
	\begin{tikzpicture}
	\draw (0,0) node{\includegraphics[width=0.9\textwidth]{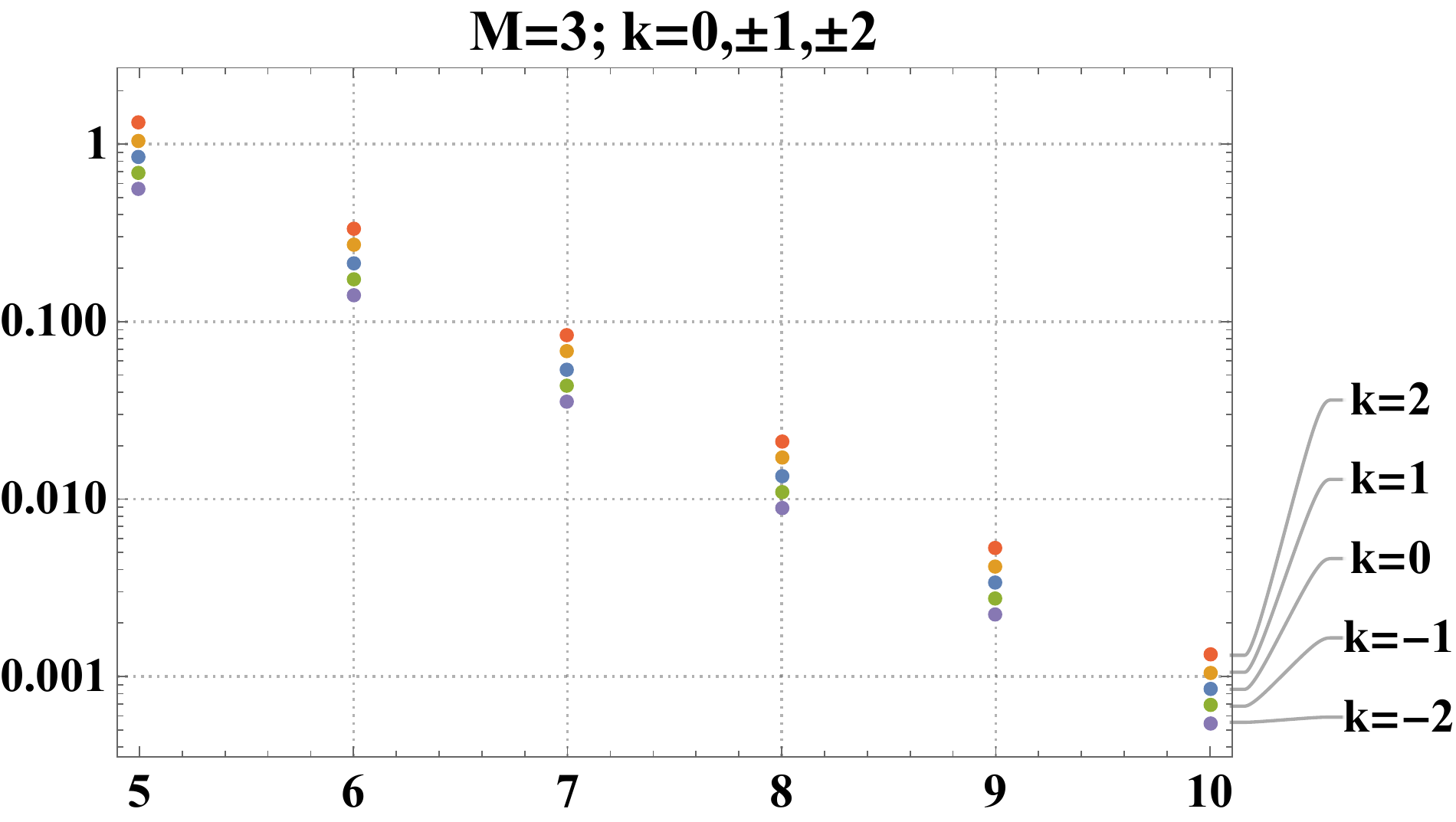}};
	\draw (5.4,-3.4) node{\textbf{N}};
	\draw (-6,3.85) node{$\mathbf{L}^{2}$-\textbf{error}};
	\end{tikzpicture}
	\caption{Upper bound to the (diagonal) $L^{2}$-norm error between the Virasoro generators $L_{k}$ and their KS approximants $L^{(N)}_{k}$ on one-particle states corresponding to the subspace of the CFT Hilbert space representing momenta up to the simulation scale $M$ as a function of the UV cutoff scale $N$. For further details, see \cite{osborneConformal2021}.}\label{fig:momrgerrork}
	\end{center}
\end{figure}

\begin{figure}
	\begin{center}
	\begin{tikzpicture}
	\draw (0,0) node{\includegraphics[width=0.9\textwidth]{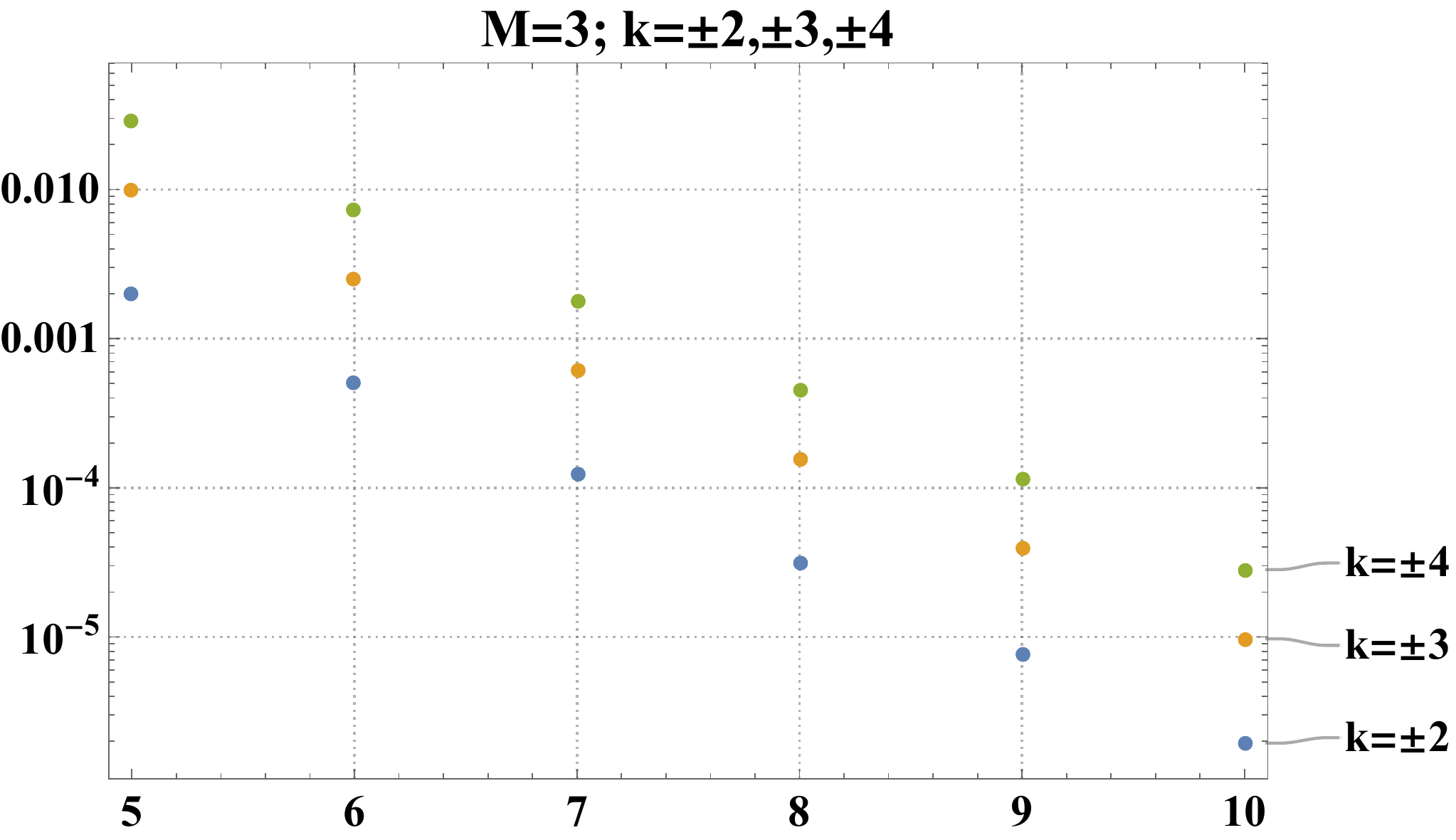}};
	\draw (5.7,-3.5) node{\textbf{N}};
	\draw (-6,3.95) node{$\mathbf{HS}$-\textbf{error}};
	\end{tikzpicture}
	\caption{Upper bound to the (off-diagonal) Hilbert-Schmidt (HS)-norm error between the Virasoro generators $L_{k}$ and their KS approximants $L^{(N)}_{k}$ on one-particle states corresponding to the subspace of the CFT Hilbert space representing momenta up to the simulation scale $M$ as a function of the UV cutoff scale $N$. Note, that this error vanishes for the generators of the Moebius group, $k=0,\pm 1$. For further details, see \cite{osborneConformal2021}.}\label{fig:momrgerrorHS}
	\end{center}
\end{figure}

\end{document}